\documentclass[namedreferences]{kluwer}

\usepackage{graphicx}
\newdisplay{guess}{Conjecture}

\begin{document}                                                                                   
\begin{article}
\begin{opening}

\title{Wind ionization structure of the short-period eclipsing LMC Wolf-Rayet binary BAT99-129: preliminary results\thanks{Based on observations obtained at the La Silla and Paranal Observatories, European Southern Observatory (Chile).}} 

\author{C. Foellmi$^1$, A.F.J. Moffat$^2$ \& S.V. Marchenko$^3$}
\institute{1. European Southern Observatory, 3107 Alonso de Cordova, 19 Vitacura, Santiago, Chile. ({\tt cfoellmi@eso.org}) \\ 2.D\'epartement de physique, Universit\'e de Montr\'eal, C.P. 6128, Succ. Centre-Ville, Montr\'eal, QC, H3C 3J7, Canada \\ 3. Department of Physics and Astronomy, Thompson Complex Central Wing, Western Kentucky University, Bowling Green, KY 42101-3576, USA.}
\runningauthor{Foellmi et al.}
\runningtitle{The short-period eclipsing LMC Wolf-Rayet binary BAT99-129.}
\date{September 30, 2005}

\begin{abstract}
BAT99-129 is a rare, short-period eclipsing Wolf-Rayet binary in the Large Magellanic Cloud. We present here medium-resolution NTT/EMMI spectra that allow us to disentangle the spectra of the two components and find the orbital parameters of the binary. We also present VLT/FORS1 spectra of this binary taken during the secondary eclipse, i.e. when the companion star passes in front of the Wolf-Rayet star. With these data we are able to extract, for the first time in absolute units for a WR+O binary, the sizes of the line emitting regions. 
\end{abstract}

\end{opening}           

\section{Presentation}  

The Wolf-Rayet (WR) star BAT99-129 (a.k.a Brey 97, see \opencite{bat99}, hereafter BAT129) has been discovered to be a short-period eclipsing binary by \inlinecite{Foellmi-etal-2003b}. It consists of a nitrogen-rich WN4 Wolf-Rayet component and a companion of unknown spectral-type. Along with BAT99-19 (Brey 16) in the LMC and HD 5980 in the SMC, BAT129 belongs to the very small group of extragalactic eclipsing Wolf-Rayet binaries. 

Absorption lines are visible in the spectrum of BAT129, especially in the blue part of the optical range ($\lesssim$4000\AA), while WR emission lines strongly dominate everywhere else. The origin of these absorption lines was unclear: they are either produced by an O-type companion, and/or they originate in a WR wind, as has been found in all the {\it single early-type} WN (WNE) stars in the SMC and, to a lesser extent, in the LMC \cite{Foellmi-etal-2003a,Foellmi-etal-2003b,Foellmi-2004}.

Previous data did not allow us to obtain the orbital motion of the companion star. Moreover, being eclipsing, the system deserved much more study. Thus, we acquired medium-resolution spectra with the EMMI spectrograph, mounted on the New Technology Telescope (NTT, La Silla Observatory, Chile). These data allowed us to resolve the orbital motion of the companion star, and find a relatively large hydrogen content on the WR star. In addition, we obtained long-slit spectra with the instrument FORS1 at the Very Large Telescope (VLT, Paranal Observatory, Chile). These spectra allow us to study the wind ionization structure of BAT129.

\section{The orbit of BAT129}  

The EMMI spectra have wavelength coverage of 3920-4380 \AA\ and a resolving power of 3500 (resolution of 2.6 pixels). The typical 30-min exposure resulted in S/N ratio between 50 and 90. The spectra were shifted to the heliocentric rest frame, and normalized to unity using continuum regions common to all spectra (see \inlinecite{Foellmi-etal-2005} for details).

We then performed an iterative method, first described in \inlinecite{Demers-etal-2002}, to reconstruct the individual mean spectra. Basically, this hypothesis-free method consists of shifting the spectra in the radial-velocity (RV) space to a common rest frame of, say, the first component and then combining them to build a high S/N spectrum of this component. This latter spectrum is shifted back to the original positions and subtracted from the original spectra. This provide a first guess of the spectrum of the second component. This procedure is started again for the second component with its respective RVs, and the new mean spectrum is again subtracted from the original spectra. This completes one full iteration. Between 3 and 5 iterations are necessary to obtain stable results. 

\begin{figure}
\centering
\includegraphics[width=0.6\linewidth]{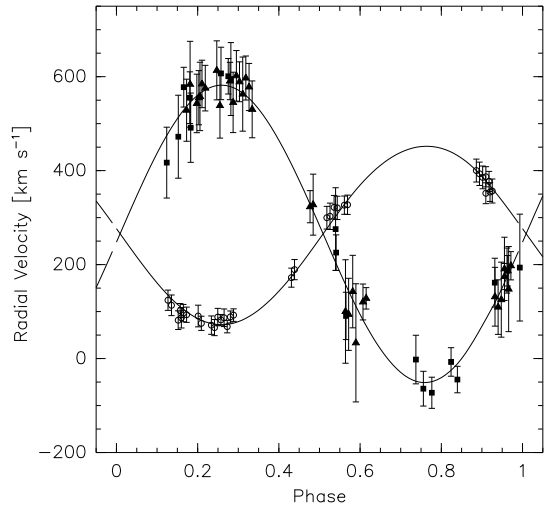}
\caption{The final orbit of the system. The orbital parameters (with their usual denominations) are the following: Period = 2.7689 $\pm$ 0.0002 days (from MACHO photometry), e=0 (assumed), K$_{WR}$ = 316 $\pm$ 5 km\,s$^{-1}$, K$_{O}$ = 193 $\pm$ 6 km\,s$^{-1}$, E$_{0}$ = 2451946.4274 $\pm$ 0.0054 (HJD), $\gamma$ = 265 $\pm$ 5 km\,s$^{-1}$.}
\label{orbit}
\end{figure}

Interestingly, a blue-shifted absorption profile is discovered in the emission lines of the WR spectrum. These absorption lines are certainly related to the WR star itself, similarly to what was found in other SMC and the LMC WR stars (see e.g. \inlinecite{Foellmi-etal-2003a}). We were then able to compute a full orbital solution (summarized in Fig.~\ref{orbit}), confirm the WN4ha spectral type of the WR star and assign an O5V spectral type to the companion \cite{Foellmi-etal-2005}. The minimum masses of the two stars are M$_{WR}$ sin$^3$~i = 14 $\pm$ 2 $M_{\odot}$ and M$_{O}$ sin$^3$~i = 23 $\pm$ 2 $M_{\odot}$, respectively, and the minimum separation $a$\,sin~$i$ = 27.9 $\pm$ 0.5 $R_{\odot}$.

\begin{figure}
\centering
\includegraphics[width=0.5\linewidth]{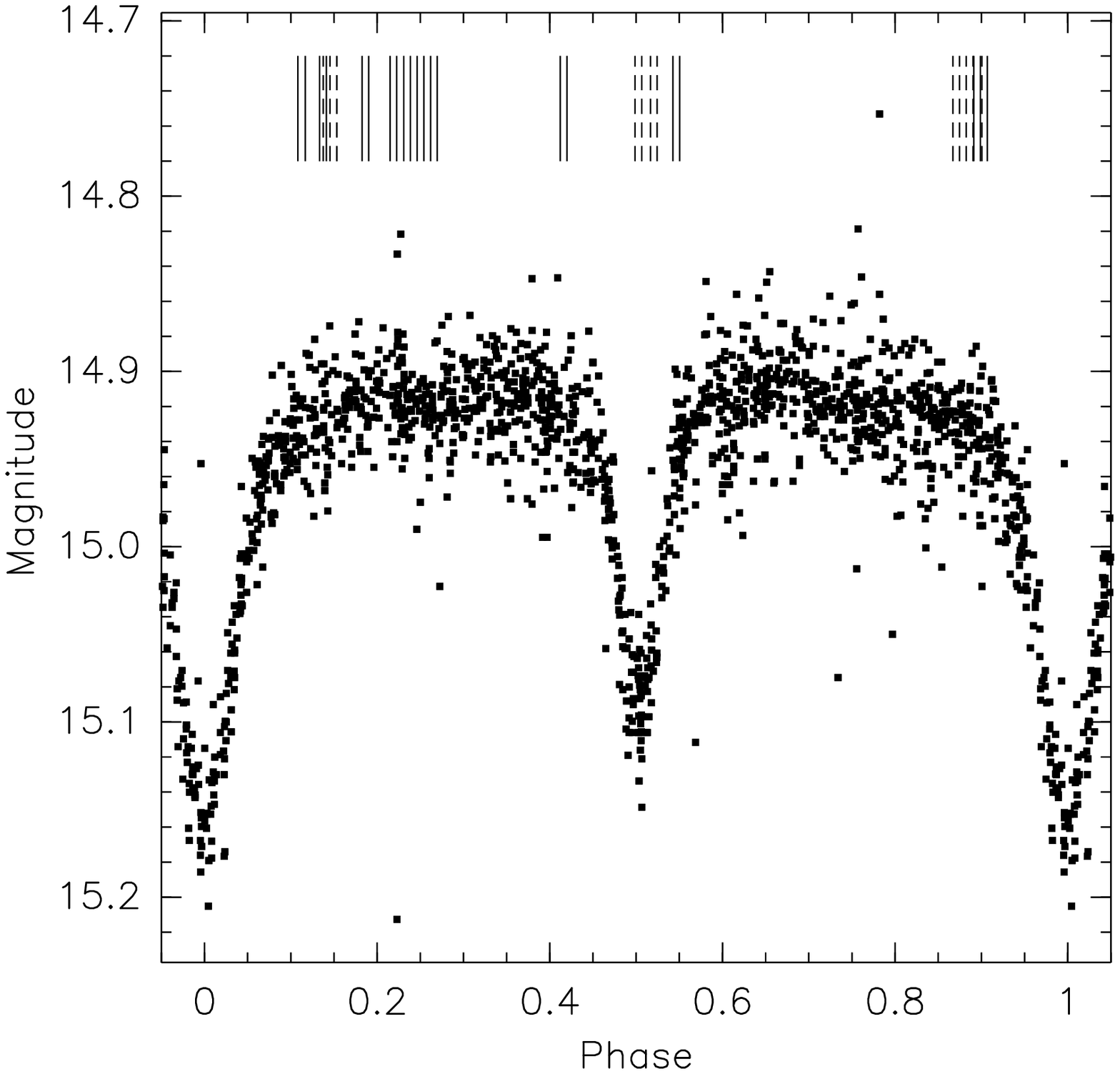}\includegraphics[width=0.5\linewidth]{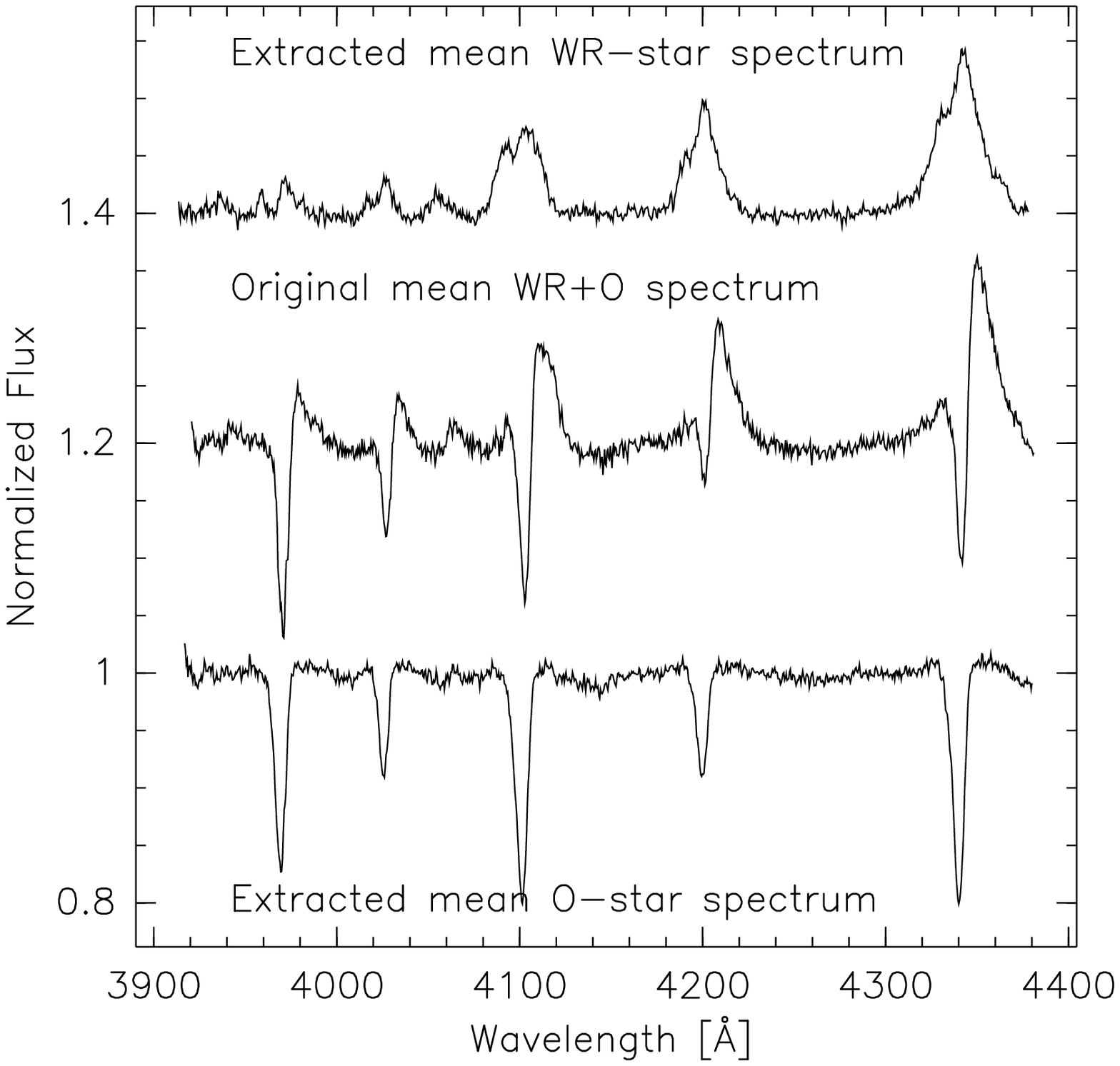}
\caption{Left panel: MACHO light curve of BAT129, as described in Foellmi et al. (2003b). The phases at which EMMI spectra were obtained are indicated by vertical lines. Right panel: result of the separation procedure. The original mean spectrum and the extracted WR spectrum have been shifted by 0.2 and 0.4 continuum units for clarity.}
\label{phot_corr}
\end{figure}

\section{The ionization structure of the wind}  

\begin{figure}
\centering
\includegraphics[width=0.6\linewidth]{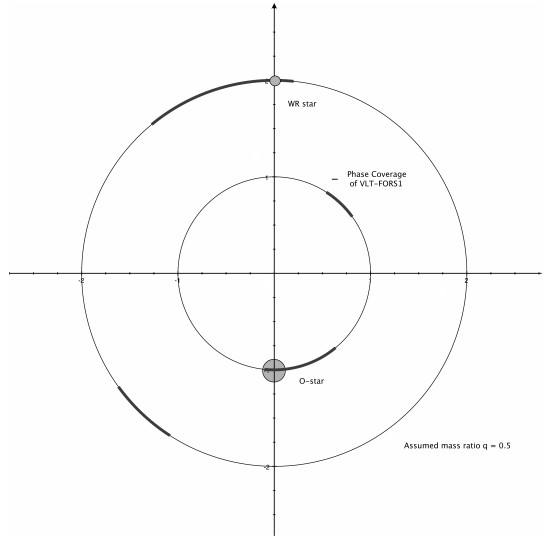}
\caption{Sketch of the observations obtained with VLT/FORS1. A mass of the O star roughly twice that of the WR star has been assumed. The two large circles indicate the respective orbital trajectories of the stars. Bold parts on these circles indicates when, according to the ephemeris of our observations, FORS1 spectra were obtained. The longest bold parts correspond to the data of the first night, while the shortest ones correspond to the second half-night.}
\label{geometry}
\end{figure}

Armed with the full set of orbital parameters of the system (except the inclination angle), we are able to use our VLT/FORS1 spectra to sketch the ionization structure of the wind of the WR component in BAT129. Our FORS1 spectra were taken in December 2003, and have a wavelength range from 3700 to 8900\AA, and a resolving power of about 500. One spectrum of S/N$\sim$200 was obtained every 6 minutes during one night through the secondary eclipse (i.e. when the O star is in front of the WR star), and identically during the next half-night. A sketch of the geometry is presented in Fig.~\ref{geometry}. It shows the orbital plane seen from above, and where the spectra were obtained during the orbit.

\begin{figure}
\centering
\includegraphics[width=0.7\linewidth]{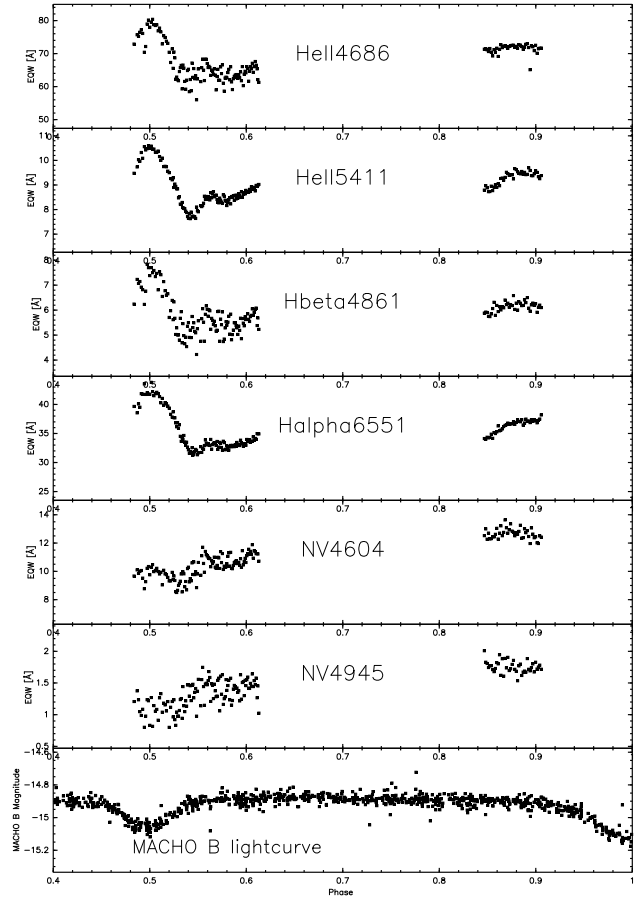}
\caption{Variations of the {\it absolute values} of the equivalent widths of various lines in the FORS1 spectra of BAT129 versus the orbital phase. The bottom panel shows the MACHO lightcurve. It can be easily seen that lines of different ions and/or different ionization potentials are not eclipsed at the same time.}
\label{eqws}
\end{figure}

The idea is as follows. We assume that in the optical the O component can be treated as a geometrically-occulting disk (i.e. a windless star); plus, the WR wind has a clear wind stratification, i.e. the line-formation zones are located at different distances from the WR core, depending on the ion and its ionization potential. Therefore, following the O star passing in front of the WR star should reveal, minute-by-minute, which region is eclipsed and during how much time. This can be measured by the variations of the equivalent widths (EWs) of the different emission lines present in the spectra (after being corrected for the varying continuum level using the MACHO lightcurve). The preliminary results are shown in Fig.~\ref{eqws}, where the normalized EWs have been plotted against the orbital phase.

\section{Discussion}  

The variations of the EWs reveal that lines of different ions and/or different ionization potentials are eclipsed at different times, {\it and with different patterns}. It is possible to roughly draw the following picture. The NV\,$\lambda$4945 line seems to follow the behavior of the photometric eclipse. This can be understood if this line is formed very close to the WR core, as does the continuum. Similarly, if we assume a perfectly symmetric pattern on each side of the photometric eclipse core (PEC), the NV\,$\lambda$4604 line has its EW eclipsed, then increased slightly at the PEC, and eclipsed again. This could be interpreted as the line is formed in a shell a bit further away from the WR core, and its projected appearance is essentially an annulus. 

The interpretation of the other lines is more difficult. At first sight, a strong increase seems to occur precisely during the photometric eclipse. However, it can be seen that the reference level, as shown in the EW values of the second half night, are slightly changing, and their mean level is {\it lower} than that of the maximum at the PEC. It is possible to interpret these variations if the regions where these lines are emitted as having an inner radius that is slightly larger than that of the O star projected radius. Therefore, the eclipsing effect of the O star is at its {\it minimum} when the O star is aligned with the WR star. When the O star moves away from this alignment, the line emitting regions start to be eclipsed. 

The situation is probably not as clear cut as this, since the orbital inclination of the system is not known. Although close to 90 degrees, it is likely that even at the PEC, the O star eclipses a (small) part of the regions where the above lines are formed. Interestingly, small wiggles in the EW variations (around phase 0.57, see Fig.~\ref{eqws}) seem to show that line emitting regions are not simple shells. These measurements have a direct impact on WR atmosphere models, such as those of \inlinecite{Hamann-Koersterke-2000}.

\theendnotes

\end{article}
\end{document}